\documentclass[aps,pra,amsmath,amssymb,amsfonts,showpacs]{revtex4}
\usepackage{graphics}
\usepackage{bm}
\usepackage{dcolumn}

\newcommand{\bmat}[3]{\bigl \langle  
       #1 \bigr| \, #2\, \bigl|     #3  
                  \bigr \rangle}

\newcommand{\ket}[1]{		\left| #1 \right>  }

\begin{document}
\title{Existential Contextuality and the Models of Meyer, Kent and
Clifton}
\author{D.M.Appleby}
\email[electronic address: ]{D.M.Appleby@qmul.ac.uk}
\affiliation{Department of Physics, Queen Mary,
University of London,  Mile End Rd, London E1 4NS,
UK}
\begin{abstract}
It is shown that the models recently proposed by Meyer, Kent and 
Clifton (MKC) exhibit a novel kind of  contextuality, which we term
existential contextuality.  It is a
particularly radical kind of contextuality,  since it is not simply
the pre-existing
\emph{value} but the actual
\emph{existence} of an observable which is context-dependent.  This
result confirms the point made elsewhere, that the MKC models do not,
as the authors claim, ``nullify'' the Kochen-Specker theorem.  It
may  also be of some independent interest.
\end{abstract}
\maketitle
\section{Introduction}

Meyer~\cite{Meyer}, Kent~\cite{KentA} and Clifton and
Kent~\cite{KentB} (MKC in the sequel)
have recently proposed a new class of hidden variables models in
which values are only assigned to a restricted subset of the set of
all observables.  MKC 
claim that their models ``nullify'' the Kochen-Specker
theorem~\cite{Bell,Koch,MerminB,Peres,Bub}.  In Appleby~\cite{self2}
we showed that this claim is unfounded:  the MKC models do not, in any
way, invalidate the essential physical point of the Kochen-Specker
theorem (for other critical discussions  see
refs.~\cite{Cabel,Havli,Mermin,self0,MKCextra}).  Nevertheless,
the MKC models are still of much interest.  Together with the models
proposed by Pitowsky~\cite{Pitowsky}   they show that the
physical interpretation of the Kochen-Specker theorem involves some
important subtleties which, in the past, have not been sufficiently
appreciated.

The purpose of this paper is to show that the MKC  models
exhibit a novel  kind of contextuality, which has not previously been
remarked in the literature, and   which is even more strikingly at
variance with classical assumptions than the usual kind of
contextuality, featuring in the Kochen-Specker theorem.   
In the usual kind of contextuality
it is only the \emph{value} assigned to an observable which is
context-dependent.  In the MKC models, however,  it is the very
\emph{existence} of an observable which is context-dependent 
(its existence,
that is, as a physical property whose value can be revealed by
measurement). This  phenomenon may be described as existential
contextuality~\footnote{It should be noted that the 
   concept of existential contextuality
   introduced here is completely unrelated to the concept of
   ontological contextuality discussed by Heywood and
   Redhead~\cite{Hey}, Redhead~\cite{Red} and Pagonis and
   Clifton~\cite{Pag}.
}. 
It confirms the point  made in
ref.~\cite{self2}, that the MKC models do not, as MKC claim, provide
a classical explanation for non-relativistic quantum mechanics.  

This paper  was
originally motivated by a seeming  inconsistency in
MKC's statement~\cite{KentA,KentB}, that their models are
both non-contextual \emph{and} 
non-local.  There do, of course, exist theories which
have both these properties (Newtonian gravity, for example). 
However, in the framework of quantum mechanics the  phenomena of
contextuality and non-locality are closely connected, as has
been stressed by Mermin~\cite{MerminB} (also see Heywood and
Redhead~\cite{Hey} and Basu
\emph{et al}~\cite{Basu}).  The discussion in Mermin~\cite{MerminB}
suggests that, if one were to examine the predictions the MKC models
make regarding (for example) the GHZ set up~\cite{GHZ}, then one
might expect to find evidence that these models  are not only
non-local (as MKC state), but also contextual (which they deny).  As
we will see, this is in fact the case.

\section{GHZ set-up, with a locality assumption}
\label{sec:loc}
\begin{figure}[htb]
\includegraphics{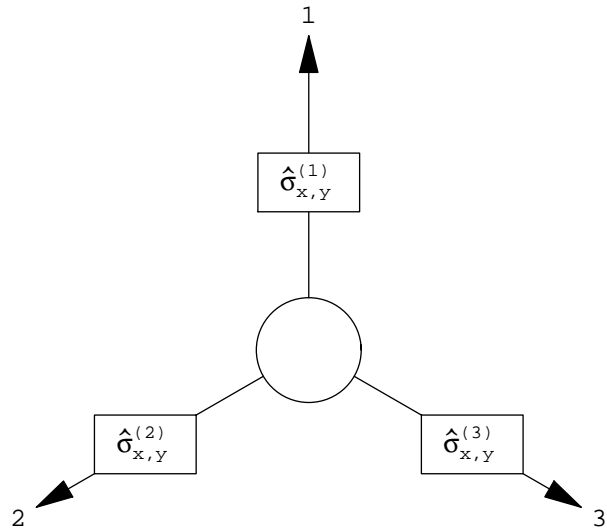}
\caption{Set-up considered in Mermin's variant of
the GHZ argument.  For each $r$, the $r^{\rm th}$
detector is set to measure one of the two target
observables
$\hat{\sigma}_{x}^{(r)}$ or
$\hat{\sigma}_{y}^{(r)}$.
\label{fig:  GHZ}}
\end{figure}

Consider  Mermin's
variant~\cite{MerminB,MerminC} of the GHZ
set-up~\cite{GHZ}, as  illustrated in
Fig.~\ref{fig:  GHZ}.   This arrangement is usually regarded as a way
of demonstrating non-locality.  As discussed in the Introduction, we
will show that it can also be used to demonstrate a form of
contextuality.

 The system 
consists of three
spin-$1/2$ particles.  Let
$\hat{\bm{\sigma}}^{(r)}$ denote the Pauli
spin vector for particle $r$, and let 
${\mathcal H}^{(r)}$ be the 2-dimensional Hilbert
space on which it acts.  The spin state is
\begin{equation}
\ket{\psi} = \frac{1}{\sqrt{2}}
\bigl(\ket{ 1,1,1 } - \ket{ -1,-1,-1}
\bigr)
\label{eq:  PsiSte}
\end{equation}
where $\ket{s_1,s_2,s_3 }$ denotes
the joint eigenstate of 
$\hat{\sigma}_{z}^{(1)}$,
$\hat{\sigma}_{z}^{(2)}$,
$\hat{\sigma}_{z}^{(3)}$
with eigenvalues
$s_1$, $s_2$, $s_3$.  The particles emerge from a
source and pass through three space-like separated
detectors (see
Fig.~\ref{fig:  GHZ}).  For each $r$ the
corresponding detector measures one of
the two  observables
$\hat{\sigma}_{x}^{(r)}$ or
$\hat{\sigma}_{y}^{(r)}$.  One
has~\cite{MerminC}
\begin{equation}
  \hat{\sigma}^{(1)}_x
  \hat{\sigma}^{(2)}_x
  \hat{\sigma}^{(3)}_x
 \ket{ \psi }
 = -\ket{\psi}
\label{eq:xxxEval}
\end{equation}
and
\begin{equation}
  \hat{\sigma}^{(1)}_x
  \hat{\sigma}^{(2)}_y
  \hat{\sigma}^{(3)}_y
  \ket{ \psi }
 =
  \hat{\sigma}^{(1)}_y
  \hat{\sigma}^{(2)}_x
  \hat{\sigma}^{(3)}_y
  \ket{ \psi }
 =
  \hat{\sigma}^{(1)}_y
  \hat{\sigma}^{(2)}_y
  \hat{\sigma}^{(3)}_x
  \ket{ \psi }
 = \ket{ \psi }
\label{eq:xyyEval}
\end{equation}
Consequently, if the
detectors are strictly ideal, and if they are
 set precisely at the combination
$xxx$, then the product of measured values must
necessarily be $-1$.  Similarly, if the
detectors are strictly ideal, and if they are 
 set precisely at one of the combinations 
$xyy$,
$yxy$,
$yyx$, then the product of measured values must
necessarily be
$+1$.

MKC argue that it would not, in practice, be possible to align the
detectors with infinite precision, implying that the detectors,
instead of performing  ideal  measurements~\footnote{It 
   should be stressed that the detectors
   still perform  \emph{non-ideal} measurements of 
   $\hat{\sigma}_{j_1}^{(1)}$, $\hat{\sigma}_{j_2}^{(2)}$,
   $\hat{\sigma}_{j_3}^{(3)}$ (see Appleby~\cite{self2}).
} 
of the observables $\hat{\sigma}_{j_1}^{(1)}$,
$\hat{\sigma}_{j_2}^{(2)}$,
   $\hat{\sigma}_{j_3}^{(3)}$  
(with $j_r= x$ or $y$),  may actually perform   ideal measurements of
a slightly different set of commuting  observables
$\hat{\tau}^{(1)}$,
$\hat{\tau}^{(2)}$,
   $\hat{\tau}^{(3)}$.  They postulate that the observables 
$\hat{\tau}^{(1)}$,
$\hat{\tau}^{(2)}$,
   $\hat{\tau}^{(3)}$ are always such that their joint
spectral resolution is a subset of a countable set  
$\mathcal{P}_{\mathrm{d}}$ (in the notation of Clifton and
Kent~\cite{KentB}), which is dense in the space of all projections on
$\mathcal{H}_1
\otimes
\mathcal{H}_2
\otimes \mathcal{H}_3$.  For each $r$ detector $r$ reveals the
pre-existing value of the observable
$\hat{\tau}^{(r)}$ which it does in fact ideally measure.

The fact that the observables $\hat{\tau}^{(r)}$ may not
precisely coincide with the observables $\hat{\sigma}_{j_r}^{(r)}$
means that there may be a  small, non-zero probability of
obtaining  the ``wrong'' measurement outcome (\emph{i.e.}\ 1 for the
combination
$xxx$, and $-1$ for the combinations $xyy$, $yxy$, $yyx$).   This
is consistent with the unavoidable imprecision of real, laboratory
measurements.

In this paper we are, for simplicity, confining  ourselves to the
kind of  measurement envisaged by MKC, in which the
imprecision is entirely due to the detectors not being aligned
precisely in the directions specified.   It should  be
stressed that such measurements  are still highly
idealised.  MKC assume that there is always \emph{some} observable
which a detector ideally measures.  They overlook the fact that a
real, laboratory instrument does  not, typically, perform an ideal
measurement of anything:  neither the nominal observable, which the
experimenter records as having been measured, nor any other
observable either.  We discuss this point further in
Appleby~\cite{self2}.

In their published papers MKC take the view  that the
difference between 
$\hat{\tau}^{(r)}$ and $\hat{\sigma}_{j_r}^{(r)}$ is  due
to  detector $r$ not being aligned with infinite precision. 
On this view
$\hat{\tau}^{(r)}$ must be a local observable of the form
$\hat{\tau}^{(r)} = \mathbf{n}_r \cdot
\hat{\bm{\sigma}}^{(r)}$, where
$\mathbf{n}_r$ is a unit vector  close to the unit
vector in the $j_r$ direction, representing the actual
alignment~\footnote{We 
  are assuming ideal detectors, so the  concept 
  ``actual alignment of detector $r$'' is unambiguous. In
  the case of non-ideal detectors this concept 
  may not be sharply defined (see
  Appleby~\cite{self2}).
 } 
of detector $r$.  Kent's~\cite{KentC} subsequent
suggestion, that 
$\hat{\tau}^{(r)}$ may be a  non-local admixture of observables
pertaining to more than one particle, will be discussed in the next
section.

Given an arbitrary  triplet of unit vectors $(\mathbf{n}_1,
\mathbf{n}_2, \mathbf{n}_3)$, define projections
\begin{equation}
\hat{P}_{s_1 s_2 s_3} = 
\frac{1}{8} (1+ s_1\, \mathbf{n}_1 \cdot \hat{\bm{\sigma}}^{(1)})
(1+ s_2\, \mathbf{n}_2 \cdot \hat{\bm{\sigma}}^{(2)})
(1+ s_3\, \mathbf{n}_3 \cdot \hat{\bm{\sigma}}^{(3)})
\end{equation}
where, for each $r$, $s_r=\pm 1$.  These projections constitute the
joint spectral resolution for the operators  $\mathbf{n}_1 \cdot
\hat{\bm{\sigma}}^{(1)}$,
$\mathbf{n}_2 \cdot \hat{\bm{\sigma}}^{(2)}$, 
$\mathbf{n}_3  \cdot \hat{\bm{\sigma}}^{(3)}$.  Define $S'_6$ to be
the set of vector triplets $(\mathbf{n}_1,
\mathbf{n}_2, \mathbf{n}_3)$ for which the corresponding 
projections $\hat{P}_{s_1 s_2 s_3}$ all 
$\in \mathcal{P}_{\mathrm{d}}$ (where, in the notation of
Clifton and Kent~\cite{KentB},
$\mathcal{P}_{\mathrm{d}}$ is the countable set of projections on
which the MKC valuations are defined). 
$S'_6$ is a countable, dense subset of
$S_2
\times S_2 \times S_2$ (where $S_2$ is the unit 2-sphere). Its
significance is that
$(\mathbf{n}_1,
\mathbf{n}_2, \mathbf{n}_3)$ represents a possible set of alignments
for the three detectors if and only if  
$(\mathbf{n}_1,
\mathbf{n}_2, \mathbf{n}_3) \in S'_6$.  

 We will now show that 
$S'_6$ cannot be a Cartesian product of the form ${S'}_2^{(1)}
\times {S'}_2^{(2)} \times {S'}_2^{(3)}$, with 
${S'}_2^{(1)},  {S'}_2^{(2)}, {S'}_2^{(3)} \subset
S^{\vphantom{(1)}}_2$.   We will then use
this to show that the MKC models exhibit a novel kind of
contextuality.

In order to establish this result suppose that $S'_6$ \emph{is}
of the form ${S'}_2^{(1)}
\times {S'}_2^{(2)} \times {S'}_2^{(3)}$.  We will show that this
assumption leads to a contradiction.

For each $r$ let $\mathbf{n}_{r x}$, $\mathbf{n}_{r y}$ be a fixed
pair of vectors $\in {S'}_{2}^{(r)}$ such that $\mathbf{n}_{rx}$
(respectively $\mathbf{n}_{ry}$) is close to $\mathbf{e}_x$
(respectvely $\mathbf{e}_y$), the unit vector in the $x$
(respectively $y$) direction.  Then
\begin{subequations}
\label{eq:nVecCond}
\begin{align}
\bmat{\psi}{ 
\bigl(\mathbf{n}_{1x} \cdot \hat{\bm{\sigma}}^{(1)}\bigr)
\bigl(\mathbf{n}_{2x} \cdot \hat{\bm{\sigma}}^{(2)}\bigr)
\bigl(\mathbf{n}_{3x} \cdot \hat{\bm{\sigma}}^{(3)}\bigr)
}{\psi} & =
-(1-\epsilon_0)
\\
\bmat{\psi}{
\bigl(\mathbf{n}_{1x} \cdot \hat{\bm{\sigma}}^{(1)}\bigr)
\bigl(\mathbf{n}_{2y} \cdot \hat{\bm{\sigma}}^{(2)}\bigr)
\bigl(\mathbf{n}_{3y} \cdot \hat{\bm{\sigma}}^{(3)}\bigr)}{\psi} & =
\phantom{-} (1-\epsilon_1)
\\
\bmat{\psi}{ 
\bigl(\mathbf{n}_{1y} \cdot \hat{\bm{\sigma}}^{(1)}\bigr)
\bigl(\mathbf{n}_{2x} \cdot \hat{\bm{\sigma}}^{(2)}\bigr)
\bigl(\mathbf{n}_{3y} \cdot \hat{\bm{\sigma}}^{(3)}\bigr)
}{\psi} & =
\phantom{-} (1-\epsilon_2)
\\
\bmat{\psi}{ 
\bigl(\mathbf{n}_{1y} \cdot \hat{\bm{\sigma}}^{(1)}\bigr)
\bigl(\mathbf{n}_{2y} \cdot \hat{\bm{\sigma}}^{(2)}\bigr)
\bigl(\mathbf{n}_{3x} \cdot \hat{\bm{\sigma}}^{(3)}\bigr)
}{\psi} & =
\phantom{-} (1-\epsilon_3)
\end{align}
\end{subequations}
where $\epsilon_a \ge 0$ for each $a$.  Let
$\epsilon = \max (\epsilon_0,\epsilon_1,\epsilon_2,\epsilon_3)$.
It follows from Eqs.~(\ref{eq:xxxEval}) and~(\ref{eq:xyyEval}) and
the continuity of the expectation values that
$\epsilon \to 0$ as $\mathbf{n}_{r x} \to \mathbf{e}_{x}$,
$\mathbf{n}_{r y} \to \mathbf{e}_{y}$ for $r=1,2,3$.  The fact that
$S'_6$ is dense in $S_2 \times S_2 \times S_2$ means that
${S'}_2^{(r)}$ is dense in $S_2$ for $r=1,2,3$.  It follows that the
vectors $\mathbf{n}_{r j}$ can be chosen so as to make $\epsilon$
arbitrarily small.

Let $\Lambda$ be the hidden state space, and for each $\lambda \in
\Lambda$ let $s_{r j}(\lambda)$ be the corresponding valuation of 
$\mathbf{n}_{r j} \cdot \hat{\bm{\sigma}}^{(r)}$.  We have $s_{r
j}(\lambda) = \pm 1$ for all $r, j$.  Define
\begin{subequations}
\begin{align} 
f_0(\lambda) & = 
- s_{1 x} (\lambda) s_{2 x} (\lambda) s_{3 x} (\lambda) \\
f_1(\lambda) & = 
\phantom{-} s_{1 x} (\lambda) s_{2 y} (\lambda) s_{3 y} (\lambda) \\
f_2(\lambda) & = 
\phantom{-} s_{1 y} (\lambda) s_{2 x} (\lambda) s_{3 y} (\lambda) \\
f_3(\lambda) & = 
\phantom{-} s_{1 y} (\lambda) s_{2 y} (\lambda) s_{3 x} (\lambda) 
\end{align}
\end{subequations}
Then $f_a (\lambda) = \pm 1$ for all $a, \lambda$.  Also
\begin{equation}
f_0 (\lambda) f_1 (\lambda) f_2 (\lambda) f_3 (\lambda)
= - \bigl( 
  s_{1 x} (\lambda)   s_{2 x} (\lambda)  s_{3 x} (\lambda)
  s_{1 y} (\lambda)   s_{2 y} (\lambda)  s_{3 y} (\lambda)
  \bigr)^2
= -1
\label{eq:fProdCond}
\end{equation}
for all $\lambda$.

Let $\mu$ be the probability measure on $\Lambda$ corresponding to
the state $\ket{\psi}$.   The assumption that 
$S'_6 = {S'}_2^{(1)} \times {S'}_2^{(2)}
\times{S'}_2^{(3)}$ implies that 
\begin{equation}
 1 -\epsilon \le 1 - \epsilon_a = \int f_a (\lambda)\, d \mu \le 1
\label{eq:fIntBound}
\end{equation}
for all $a$.  For each $a$ let $A_a$ be the set
\begin{equation}
A_a = \{\lambda\in \Lambda\colon f_a(\lambda) = 1\}
\end{equation}
Then it follows from Inequality~(\ref{eq:fIntBound}) that
\begin{equation}
  1 - \epsilon \le \int f_a(\lambda)\, d \mu = 2 \mu(A_a) -1
\end{equation}
for all $a$.  It follows that $\mu(A_a) \ge 1 - \epsilon/2$ for all
$a$ and, consequently, that
$\mu(A_0 \cap A_1 \cap A_2 \cap A_3)  \ge 1-  2\epsilon$.  We noted
above that, with a suitable choice of the vectors $\mathbf{n}_{r j}$,
$\epsilon$ can be made arbitrarily small.  It follows that there
exist vectors $\mathbf{n}_{r j}$ such that 
$\mu(A_0 \cap A_1 \cap A_2 \cap A_3)>0$ (in fact, there exist vectors
$\mathbf{n}_{r j}$ such that $\mu(A_0 \cap A_1 \cap A_2 \cap A_3)
\approx 1$).  On the other hand, it follows from
Eq.~(\ref{eq:fProdCond}) that
$\mu(A_0 \cap A_1 \cap A_2 \cap A_3) = 0$ for every choice of
$\mathbf{n}_{r j}$---which is a contradiction.

We have thus shown that the set $S'_6$ does not have the form of a
Cartesian product for any model of MKC type.  This has important
consequences:  for it implies that it must, in
general, happen that  a change in the alignment of one detector forces
a change in the alignment of at least one of the other two detectors. 
This represents a form of non-locality.   However, the point which
concerns us here is that it also represents a form of contextuality.

It is a particularly striking form of contextuality.   Let
${S'}_2^{(r)}$ be the set of possible alignments for detector $r$. 
In the usual kind of contextuality  ${S'}_2^{(r)}$ is fixed,
and  it is only the values assigned to the members of this set
which depend on the measurement context.  However, in the MKC models
it is the set 
${S'}_2^{(r)}$ itself which depends on the measurement context.  In
other words, it is not simply the  \emph{value},
but the very \emph{existence}  of an observable which is
context-dependent (its existence, that is, as  a physical property
whose value can be revealed by measurement).

\section{GHZ set-up, with non-local detectors}
\label{sec:nonloc}
In the last section we assumed that detector $r$ reveals the value of
a local observable, defined on the state space of particle $r$. 
Kent~\cite{KentC} has objected to this assumption.  He suggests,
instead, that, on  the level of the hidden variables, the detectors
may function as non-local devices,  which reveal the values of
non-local admixtures of observables pertaining to more than one
particle.

Let us begin by noting that this suggestion involves a significant
departure from the view taken in MKC's published papers.  In their
published papers MKC argue that the observable
$\hat{\tau}^{(r)}$, whose value is revealed by detector $r$,  is also
the observable which detector
$r$  ideally 
measures~\footnote{MKC
   only consider ideal detectors. Their assumption, that a measurement
   reveals the pre-existing value
   of the observable which is ideally measured, is obviously not 
    applicable in the case of non-ideal detectors (see
   Appleby~\cite{self2}). 
}, 
the discrepancy between
$\hat{\tau}^{(r)}$ and $\hat{\sigma}_{j_r}^{(r)}$ being entirely
attributable to an inaccuracy in the alignment of detector $r$. 
Clearly, detector $r$ can only perform ideal quantum measurements of
local observables pertaining to particle $r$.  Consequently, the
position adopted in MKC's published papers implies that
$\hat{\tau}^{(r)}$ must be a local observable pertaining to particle
$r$---as we assumed in the last section.

If a detector  reveals the pre-existing value of some non-local
observable then, on the level of the hidden variables, it must be
interacting non-locally with more than one particle.  This
interaction would represent a further element of non-locality in the
theory, additional to the non-locality required by the standard
arguments (Bell, GHZ, \emph{etc}.).  A model of this kind  would thus
be even more strongly non-classical than the models originally
proposed in MKC's published papers.

Nevertheless, the fact that the observables $\hat{\tau}^{(r)}$ may be
assumed to be arbitrarily close to local observables of the form
$\mathbf{n}_r \cdot \hat{\bm{\sigma}}^{(r)}$ means that a model of the
kind indicated will still be consistent with the empirical
predictions of conventional quantum mechanics.  The question
consequently arises, whether the phenomenon of existential
contextuality, discussed in the last section, also occurs in models
of this more general kind.  It is easily seen that the answer to this
question is in the affirmative.

Let $\mathcal{P}$ be the set of all projection operators
on $\mathcal{H}_1\otimes\mathcal{H}_2\otimes\mathcal{H}_3$, and let
$\mathcal{P}_{\mathrm{d}}$ be the countable, dense subset of
$\mathcal{P}$ on which the MKC truth-functions are defined
(where we are employing the notation of Clifton and
Kent~\cite{KentB}, as before).   Let
$\bar{\mathcal{P}}_{\mathrm{d}}$ be the set of  self-adjoint
operators on $\mathcal{H}_1\otimes\mathcal{H}_2\otimes\mathcal{H}_3$
whose spectral resolutions are contained in $\mathcal{P}_{\mathrm{d}}$.   The triplet  
$(\hat{\tau}^{(1)}, \hat{\tau}^{(2)},
\hat{\tau}^{(3)})$ of commuting observables whose values are revealed
by the three detectors must
$\in \bar{\mathcal{P}}_{\mathrm{d}}\times
\bar{\mathcal{P}}_{\mathrm{d}}\times
\bar{\mathcal{P}}_{\mathrm{d}}$.  It is determined by the hidden state
of the three detectors.  Let $T \subseteq 
\bar{\mathcal{P}}_{\mathrm{d}}\times
\bar{\mathcal{P}}_{\mathrm{d}}\times
\bar{\mathcal{P}}_{\mathrm{d}}$ be the set of all possible triplets
$(\hat{\tau}^{(1)}, \hat{\tau}^{(2)},
\hat{\tau}^{(3)})$, as determined by the set of all possible hidden
detector states.  The set $T$ is  the analogue, in the more
general setting of this section, of the set $S'_6$ defined in the
last section.

We may now show, using a straightforward  modification of the
argument in the last section, that $T$ is not a Cartesian product. 
In fact, suppose that $T$ \emph{was} of the form $T^{(1)} \times
T^{(2)}
\times T^{(3)}$ (with
$T^{(r)}
\subseteq \bar{\mathcal{P}}_{\mathrm{d}}$ for $r=1,2,3$).  We could
then choose, for each $r=1,2,3$ and $j= x,y$, operators
$\hat{\tau}_{j}^{(r)} \in T^{(r)}$ such that
$\hat{\tau}_{j}^{(r)} \approx
\hat{\sigma}_{j}^{(r)}$ for all $r$, $j$.  This would imply
\begin{subequations}
\begin{align}
1-\epsilon & 
   \le - \bmat{\psi}{\hat{\tau}_{x}^{(1)}
   \hat{\tau}_{x}^{(2)}\hat{\tau}_{x}^{(3)}}{\psi}
   \le 1
\\
1-\epsilon & 
   \le \phantom{-} \bmat{\psi}{\hat{\tau}_{x}^{(1)}
   \hat{\tau}_{y}^{(2)}\hat{\tau}_{y}^{(3)}}{\psi}
   \le 1
\\
1-\epsilon & 
   \le \phantom{-} \bmat{\psi}{\hat{\tau}_{y}^{(1)}
   \hat{\tau}_{x}^{(2)}\hat{\tau}_{y}^{(3)}}{\psi}
   \le 1
\\
1-\epsilon & 
   \le \phantom{-} \bmat{\psi}{\hat{\tau}_{y}^{(1)}
   \hat{\tau}_{y}^{(2)}\hat{\tau}_{x}^{(3)}}{\psi}
   \le 1
\end{align}
\end{subequations}
where the positive constant $\epsilon$  can be chosen arbitrarily
small (compare 
Eqs.~(\ref{eq:nVecCond}) in the last section).  If $\epsilon < 1/2$
we  can show that these inequalities lead to a contradiction, by an
argument which is essentially the same as the argument following 
Eqs.~(\ref{eq:nVecCond}) in the last section. It follows that $T$ is
not a Cartesian product.

We conclude that the MKC models
still exhibit the phenomenon of existential contextuality described
in the last section, even on the assumption that the detectors may
reveal the pre-existing values of non-local observables.

\section{Conclusion}
\label{sec:conclusion}

In this paper we have argued
that the MKC models are contextual.  It follows that they do
not, as MKC claim, provide a classical explanation for the
empirically verifiable predictions of non-relativistic quantum
mechanics.  This confirms the conclusion reached in
Appleby~\cite{self2}, on the basis of a  different, completely
independent argument.

We would, however, stress that, notwithstanding these criticisms,
it appears to us that
the work of MKC is deeply interesting, and important. 
We have argued that MKC's attempt to explain non-relativistic
quantum mechanics in classical terms is misconceived.  Nevertheless
their work is still valuable because, together with the
earlier work of Pitowsky~\cite{Pitowsky}, it shows that the physical
interpretation of the Kochen-Specker theorem is a great deal more
subtle than  may superficially appear.  It consequently leads to a
 deeper understanding of the conceptual implications of
quantum mechanics.

The work of MKC and Pitowsky is also interesting because it
  enlarges the scope of the hidden variables concept in a new and
imaginative way. In the past hidden variables theories have
primarily been motivated by purely philosophical considerations.   
The emphasis has been largely (though not entirely---see
Valentini~\cite{Val1,Val2}) on constructing alternative
interpretations of conventional quantum mechanics. Recently, however,
't~Hooft~\cite{tHooft1,tHooft2} (in one way) and Faraggi and
Matone~\cite{Far1} and Bertoldi~\emph{et al}~\cite{Far2} (in another
way) have speculated that Planck scale physics may most appropriately
be described in terms of a hidden variables theory which is
\emph{not}  equivalent to conventional quantum 
mechanics.  A theory of this kind, if it could be constructed, would
be \emph{empirically} significant.

In this connection it may be worth noting 
that 't~Hooft~\cite{tHooft2}
has argued that, on the level of Planck scale variables,
it may not be possible to rotate a detector at will
so as to measure either the $x$ or $y$ components of
a particle's spin, 
and that this may
provide a way of circumventing the Bell theorem.
This  proposal is  similar to MKC's   attempt to
circumvent the Kochen-Specker theorem.  Our analysis of
the MKC models would consequently seem to indicate that one cannot
restore classicality in the manner 't~Hooft suggests. 

On the other hand, there is no evident reason why one should demand
non-contextuality and locality in respect of  a theory of the kind
proposed by 't~Hooft.   Such a theory must, by definition,
restore the concept  of a world of objective facts.   However, this
concept is by no means exclusive to classical physics.  In other
respects the theory might be highly non-classical.  Indeed, it might
be even more highly non-classical than conventional quantum
mechanics.  
The aim  is to understand the actual constitution of
the physical universe.  There is no clear reason
to exclude, at the outset, the possibility that the world actually
is contextual and non-local.  

The ideas of 't~Hooft and Faraggi~\emph{et al} are admittedly
speculative.  They do, however, provide an additional motive for
investigating new and more imaginative implementations of the hidden
variables hypothesis.

\begin{acknowledgments}
It is a pleasure to thank Prof.\ A.~Zeilinger for his hospitality
during the programme ``Quantum Measurement and Information'' held
at ESI in Vienna.  It is also a pleasure to thank 
N.D.~Mermin,  A.~Peres,  A.~Kent,  P.~Busch, D.~Home, K.~Svozil,
I.~Pitowsky, G.~Mahler, and an anonymous referee for their
stimulating and helpful comments.
\end{acknowledgments}

\end{document}